\documentclass[10pt,prd,twocolumn,preprintnumbers,amsmath,amssymb,superscriptaddress]{revtex4-1}

\usepackage{import}
\usepackage{hyphenat}
\usepackage{array}
\usepackage[british]{babel}
\usepackage[]{graphicx}
\usepackage{color}			
\usepackage{pict2e}			
\usepackage{empheq}
\usepackage{bbm}
\usepackage[colorlinks]{hyperref}
\usepackage{braket}

\begin{document}
\title{A model with two periods of inflation}
\author{Simon Schettler}
\affiliation{Institut f\"ur Theoretische Physik, Universit\"at Heidelberg\\ 
Philosphenweg 16, D-69120 Heidelberg, Germany}
\author{J\"urgen Schaffner-Bielich}
\affiliation{Institut f\"ur Theoretische Physik, Goethe-Universit\"at,
Max-von-Laue-Stra\ss e 1, D-60438 Frankfurt, Germany}
\date{\today}

\begin{abstract}    
A scenario with two subsequent periods of inflationary expansion in the very early universe is examined. 
The model is based on a potential motivated by symmetries being found in field theory at high energy. 
For various parameter sets of the potential the spectra of scalar and tensor perturbations that are expected to originate from this scenario are calculated. 
Also the beginning of the reheating epoch connecting the second inflation with 
thermal equilibrium is studied. 
Perturbations with wavelengths leaving the horizon around the transition between the two inflations are special: 
It is demonstrated that the power spectrum at such scales deviates significantly from expectations based on measurements of the cosmic microwave background (CMB). 
This supports the conclusion that parameters for which this part of the spectrum leaves observable traces in the CMB must be excluded. 
Parameters entailing a very efficient second inflation correspond to standard small-field inflation and can meet observational constraints. 
Particular attention is paid to the case where the second inflation leads solely to a shift of the observable spectrum from the first inflation. 
A viable scenario requires this shift to be small.
\end{abstract}

\maketitle

\section{Introduction}
Cosmological inflation is a time period with accelerated growth of the scale parameter $a(t)$. If it is included into the cosmic history before the hot Big Bang evolution, the inflationary scenario poses a significant alleviation to several problems: For example observational results like the small spatial curvature at large scales and the isotropy of the CMB are a natural outcome of early universe inflation \cite{Guth81}. A further reason to consider inflation is that it provides a mechanism for the production of primordial inhomogeneities out of which the structures of the later universe can form \cite{MukhanovFB92}.

A large class of solutions with accelerated scale parameter relies on slowly evolving scalar fields. This class is called slow-roll inflation. By definition, the time derivatives of the field values meet certain slow-roll conditions. They can also be formulated as flatness conditions of the potential. A great variety of potentials of scalar fields fulfill these requirements and can lead to inflation. Many potentials have been suggested as the origin of inflation \cite{Guth81,Linde82,Linde83,Linde91,RandallSG96}, and measurements have been compared to the corresponding predictions \cite{HawkingMS82,Planck15}.  
In the following, observational consequences of two subsequent periods of inflation are considered.
A similar scenario has also been discussed in Refs.~\cite{JainCGSS09,JainCSS10} and early discussions of double inflation can be found in Refs.~\cite{KofmanLS85,SilkT87}. References~\cite{KanazawaKSY00,YamaguchiY01} contain discussions of double inflation in the context of supergravity. The production of black holes and cosmic strings in similar scenarios is studied in Refs.~\cite{GarciaBellidoLW96,KawasakiSY98,Yokoyama98,Linde13}.
Two inflations can occur for potentials being flat in two regions of field space. The potential used for our calculations is discussed in the next section. Section \ref{s:Homogeneous} is concerned with the evolution of a homogeneous scalar field within this potential. Section \ref{s:Fluctuations} describes the fluctuations resulting from the two inflations. 
The decay of the inflaton field after inflation is the subject of Section \ref{s:Preheating} which is followed by some conclusions in Section \ref{s:Conclusion}.

\section{The potential and its simplification}

The inflationary potential used in our calculations originally stems from effective theories of the strong interaction. The goal of these theories is to describe strongly interacting matter on a basis simpler than the theory of quantum chromodynamics (QCD). The properties of QCD, especially its symmetries, are used as a guideline to construct the Lagrangian of the effective theory in question. A prominent example is the linear sigma model \cite{Gell-MannL60} which incorporates chiral symmetry in terms of effective pion and sigma fields $\vec\pi$ and $\sigma$. There are various extensions of this model which are constructed as to reflect further characteristics of QCD. One example is the inclusion of a dilaton field $\chi$ which controls the behavior of the terms of the Lagrangian under dilatations \cite{BoeckelS10,CampbellEO90,MishustinBR93,PapazoglouEa97}. In a simple form, the potential of the theory then reads
\begin{equation}
\begin{split}
	V = \; &\frac{\lambda}{4}\left( \vec\pi^2 + \sigma^2 \right)^2
	- v_0^2 \,( \vec\pi^2 + \sigma^2 )\left(\frac{\chi}{v} \right)^2
	\\ &- f_\pi m_\pi^2 \sigma \left(\frac{\chi}{v} \right)^2
	+ k \left(\frac{\chi}{v}\right)^4 + \frac{1}{4} \chi^4 \ln\left(\frac{\chi^4}{v^4}\right),
\end{split}
\label{eq:Potential}
\end{equation}
where certain constants $\lambda$, $v_0$, and $v$ have been defined. 
The second term leads to spontaneous breaking of chiral symmetry. Explicit symmetry breaking is introduced by the third term leading to massive pions. It is proportional to the pion decay constant $f_{\pi}$ and the squared pion mass $m_{\pi}^2$.  The last two terms are the dilaton potential with minimum at $\chi=v$. The factors of $\chi$ in the first three terms yield the expected transformation behavior under scale transformations or dilatations, $x^\mu \rightarrow a x^\mu$, $a\neq 0$, see Ref.~\cite{Coleman85}.

\begin{figure}
	\centering
	\includegraphics[trim=30 15 30 30,scale=0.65]{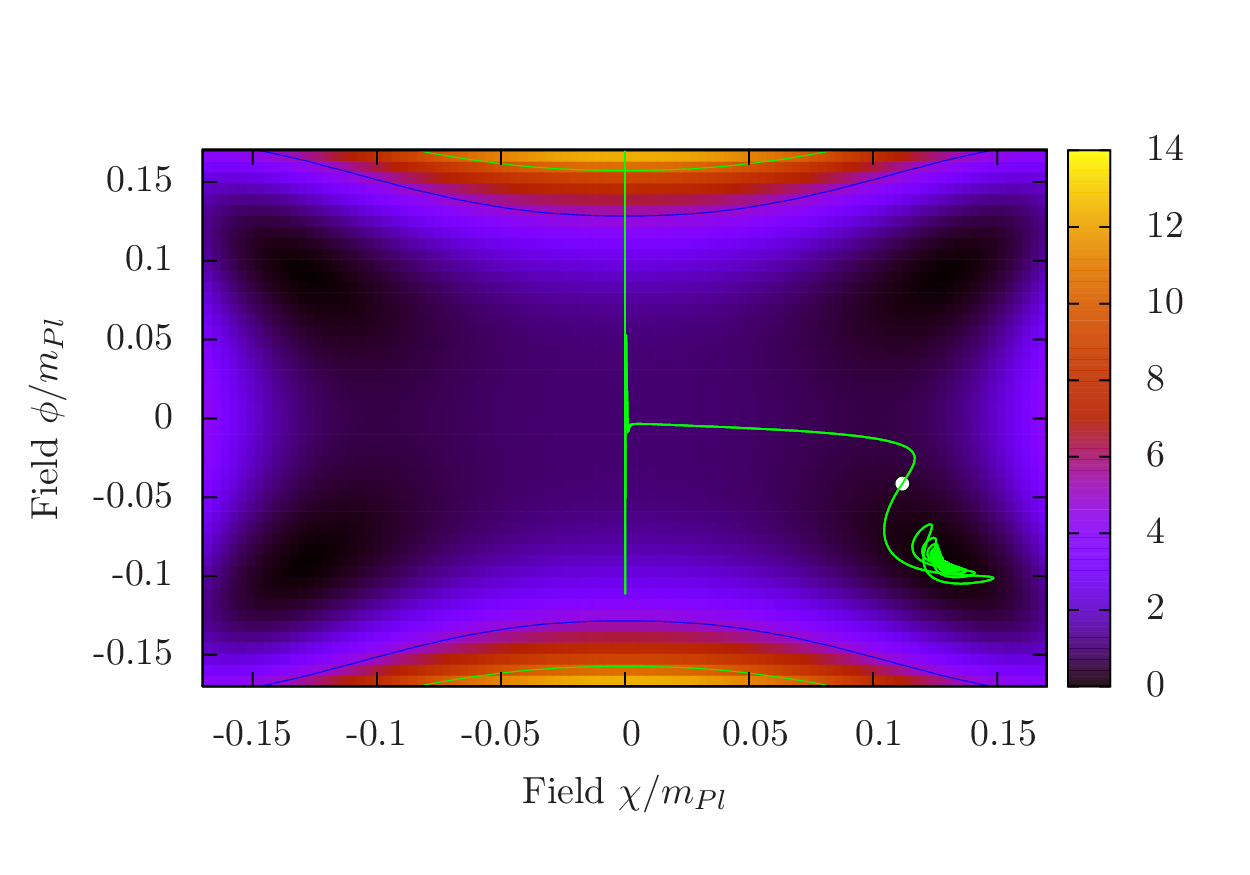} 
	\caption{Possible way of the fields $\phi$ and $\chi$ through the potential Eq.~\eqref{eq:FullPotPhiChi}.  In this case the energy scale and the couplings are chosen as $v=0.1m_\text{Pl}$, $\lambda_0=\lambda_1=\lambda_2=10^{-12}$. The white dot marks the end of accelerated expansion. The color of the background encodes the value of the potential, $V(\phi,\chi)/V_0$.}
\label{fig:t401_PotWithPath}	
\end{figure}

The potential Eq.~\eqref{eq:Potential} is motivated for use in high energy physics by its high degree of symmetry and its construction after the example of QCD. For our application to inflation the potential has been simplified: 
The third, symmetry breaking term has been omitted and the chiral fields $\vec\pi$ and $\sigma$ have been reduced to one, $\phi$. 
 So the calculations have been done on the chiral circle. 
Renaming of the constants then leaves
\begin{equation}
	V = V_0 + \frac{1}{4} \lambda_0 \chi^4 \left(\ln \left|\frac{\chi}{v}\right|-\frac{1}{4}\right) + \frac{1}{4} \lambda_1 \phi^4 - \frac{1}{4} \lambda_2 \chi^2 \phi^2,
\label{eq:FullPotPhiChi}
\end{equation}
where the offset
\begin{equation}
	V_0= \frac{1}{16}\lambda_0 v^4 \, \text{exp}{\left(\lambda^2_2/\lambda_0\lambda_1\right)}
\label{eq:V_0ForFullPot}
\end{equation}
keeps the potential positive. 
The potential Eq.~\eqref{eq:FullPotPhiChi} is displayed in Fig.~\ref{fig:t401_PotWithPath} along with a possible path $(\chi(t),\phi(t))$ during and after inflation. The color code reflects the value of the potential $V(\chi,\phi)$ in units of $V_0$. The constants are chosen as $\lambda_0=\lambda_1=\lambda_2=10^{-14}$ and $v=m_\text{Pl}/10$ and the calculation of the path is done without consideration of field inhomogeneities. The starting point of the fields is at large $\phi$ and small positive $\chi$ with zero initial field velocities. Then $\phi(t)$ starts to evolve down the quartic potential while the value of $\chi$ is still close to zero, i.e. much smaller than the symmetry breaking scale $v$. During this period of time the evolution can be described as large field inflation with an offset $V_0$ in the potential. After one and a half oscillations the field $\phi$ stays close to zero due to Hubble damping. Now the main evolution in field space is a slow roll along the $\chi$ direction. Accordingly, this period of time can be characterized as a small-field hilltop inflation. Between these two inflationary stages the universe undergoes a short phase of decelerated expansion. 

This model is related to the standard potential for hybrid inflation introduced in~Ref.~\cite{Linde91}. One difference is that in the potential Eq.~\eqref{eq:FullPotPhiChi} there is no channel the field $\chi$ could be confined to during the first inflationary period. In order not to terminate inflation too early, the initial values of $\chi$ and $\dot\chi$ therefore have to be very close to zero, as mentioned above. 
%
%
%
%
%
%
%
Then the field $\chi$ does not act as a waterfall field that rapidly rolls down to the minimum of the potential after a critical value of $\phi$ has been passed. Instead, it  could lead to a second period of inflation.

This second inflation is due to the symmetry breaking logarithmic term in the potential. If $\lambda_2 \neq 0$, the last term in Eq.~\eqref{eq:FullPotPhiChi} also contributes. It comes from spontaneous chiral symmetry breaking in the linear sigma model. Because it can considerably enlarge the offset $V_0$, Eq.~\eqref{eq:V_0ForFullPot}, the inflationary range in field space is extended to $\chi \approx v$ for large enough $\lambda_2$. The scenario ends with oscillations of the scalar fields around one of the minima in Fig.~\ref{fig:t401_PotWithPath}. The particle production that accompanies the oscillations reheats the universe and the standard hot Big Bang history sets in. 

In the rest of this section a simplification of the potential Eq.~\eqref{eq:FullPotPhiChi} will be discussed. To justify this simplification it first needs to be stated that fluctuations at a given length scale are generated when the growing wavelength is of the same order as the Hubble length $1/H\equiv(\dot a/a)^{-1}$. This is also called the time of horizon crossing or horizon exit of the mode.

The potential discussed above includes two scalar fields both acting as inflaton fields which by definition drive the inflation. This means that the resulting fluctuations should in general be calculated within the framework of multifield inflation \cite{GordonWBM00, Hwang91, Wands07}. Then one has to care for adiabatic modes which correspond to fluctuations along the inflaton field and for entropy modes characterizing fluctuations between the energy contributions of different fields without changing the total energy density $\rho$.
Also after horizon exit, entropy modes can source adiabatic fluctuations with equal wavelengths. 
As discussed in Ref.~\cite{GordonWBM00} however, entropy fluctuations are only produced when the background fields follow a curved trajectory in field space.
Thus, only during the time when the main field evolution changes its direction from $\pm \phi$ to $\chi$, the development of entropy fluctuations is expected. 
But since at this stage the field $\phi$ is not evolving slowly anymore and no accelerated expansion takes place, there will be no production of non-adiabatic fluctuations even in this moment. 
So within this scenario no complication due to multifield dynamics will occur.    
Therefore, the calculations can be done within the simplified setting of single field inflation.

\begin{figure}
	\centering
	\includegraphics[scale=0.65]{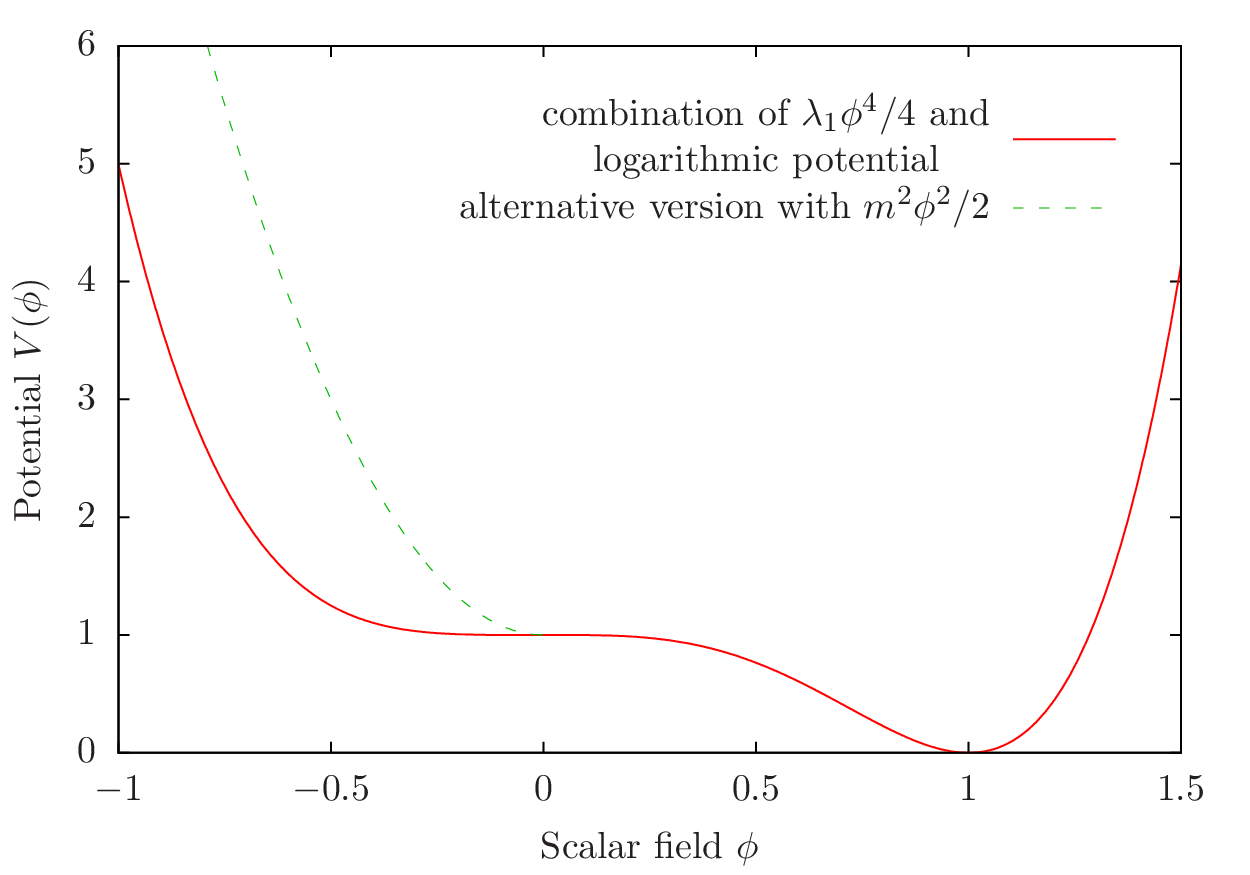} 
	\caption{The two potentials Eqs.~\eqref{eq:SimplPotPhi4} and \eqref{eq:SimplPotPhi2} that are used in the calculations of fluctuations produced during inflation.}
\label{fig:t402b_PotForCalcs_bw}	
\end{figure}

The two inflationary epochs are both ascribed to the same scalar field $\phi$ evolving within an adapted potential. 
More precisely the computation of fluctuations is done using  the two alternative potentials    
\begin{equation}
	V_4(\phi) = 
	\begin{cases}
		V_0 + \frac{1}{4} \lambda_1 \phi^4 & \phi < 0\\
		V_0 + \frac{1}{4} \lambda_0 \phi^4 \left(\ln \left|\frac{\chi}{v}\right|-\frac{1}{4}\right) & \phi \ge 0
	\end{cases}
\label{eq:SimplPotPhi4}
\end{equation}
and
\begin{equation}
	V_2(\phi) = 
	\begin{cases}
		V_0 + \frac{1}{2} m^2 \phi^2 & \phi < 0\\
		V_0 + \frac{1}{4} \lambda_0 \phi^4 \left(\ln \left|\frac{\chi}{v}\right|-\frac{1}{4}\right) & \phi \ge 0,
	\end{cases}
\label{eq:SimplPotPhi2}
\end{equation}
see Fig.~\ref{fig:t402b_PotForCalcs_bw}. In both cases $V_0$ is given by Eq.~\eqref{eq:V_0ForFullPot}.

\section{The homogeneous mode}\label{s:Homogeneous}
\begin{figure}
	\centering
	\includegraphics[scale=0.65]{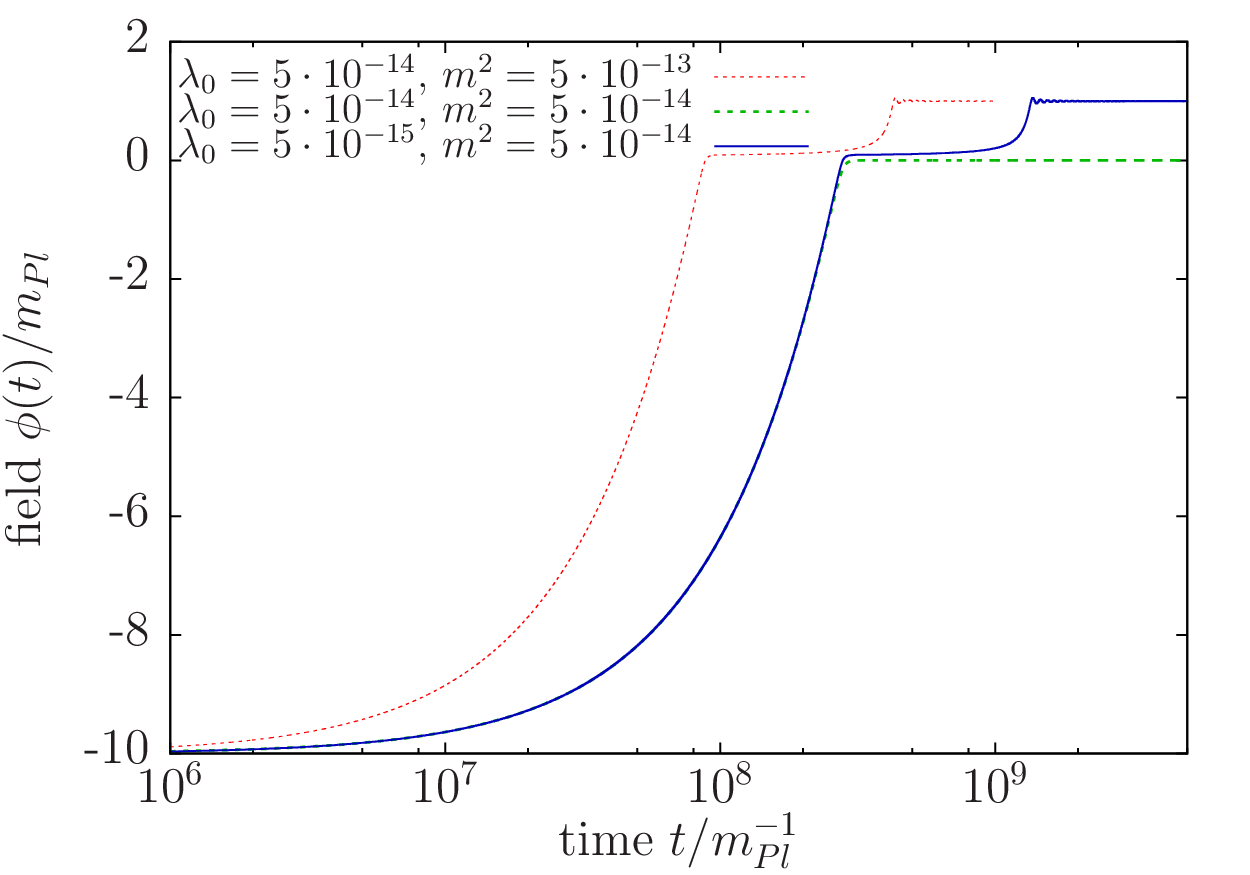} 
	\caption{The scalar field $\phi$ rolling down the potential Eq.~\eqref{eq:SimplPotPhi2} for three different parameter sets with $v=1m_\text{Pl}$ in each case. Too large $\lambda_0/m^2$ retains the field for a very long time at values around $\phi=0$ (green curve).}
\label{fig:t410b_QuadDilFields_6-7-10_bw}	
\end{figure}

\begin{figure}
	\centering
	\includegraphics[scale=0.65]{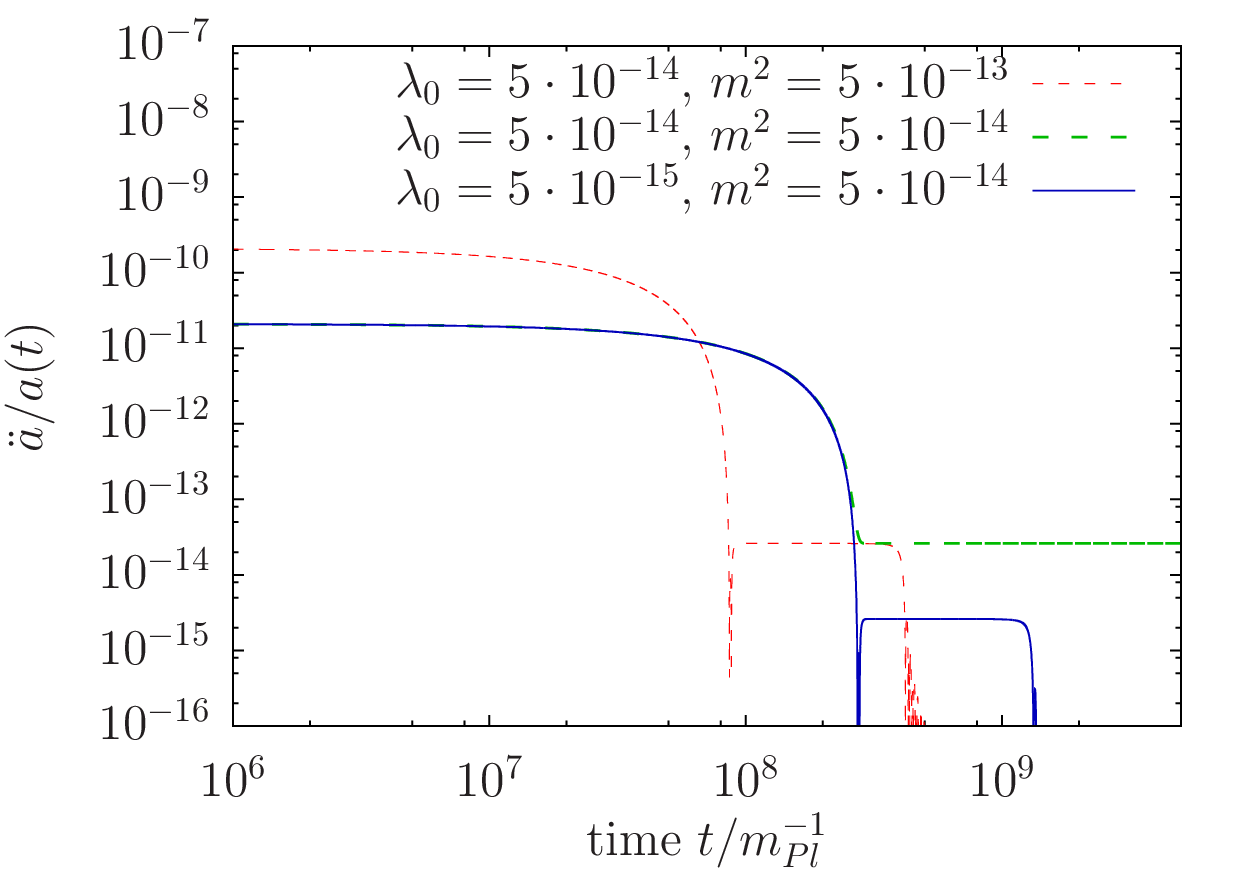} 
	\caption{Absolute value of the acceleration of the scale parameter $a(t)$ for the same three parameter sets as in Fig.~\ref{fig:t410b_QuadDilFields_6-7-10_bw}. During the steep dips in the cases one and three, $\ddot a$ changes its sign twice. During these short intervals, inflation stops and the expansion of the universe is decelerated. The green line shows no dip, which corresponds to a single, uninterrupted inflation.}
\label{fig:t412_QuadDiladdot_6-7-10}	
\end{figure}

Within the simplified scenario described in the last section, the field starts at a super--Planckian negative value and moves slowly down to the flat region around $\phi=0$. 
In spite of vanishing potential gradient, for small field values slow roll ends for a short time within many parameter sets. 
The evolution at $\phi > 0$ uses the same hilltop potential for both variants $V_2$ and $V_4$. 
%
%
The solution $\phi(t)$ is sketched in Fig.~\ref{fig:t410b_QuadDilFields_6-7-10_bw} for three different sets of couplings and the same symmetry breaking scale $v=1m_\text{Pl}$ in each case. For most of the evolution of the homogeneous mode down to $\phi =0$, the contribution of the offset $V_0$ is negligible. 
For the parameter set $\lambda_0 = 5\cdot 10^{-14}$, $m^2 = 5\cdot 10^{-14}m_\text{Pl}^2$, the evolution of $\phi(t)$ is stopped at $\phi = 0$. This happens when the inflaton is slowly rolling until it reaches this value. Due to the negligible kinetic energy, $\phi$ is stopped by the Hubble drag and does not overcome the region with $V'\approx 0$. 
Corresponding phenomena are found in the field motion within the potential Eq.~\eqref{eq:SimplPotPhi4}, and are not displayed. This is also the case with the content of the following figures. 
%
%
Fig.~\ref{fig:t412_QuadDiladdot_6-7-10} shows the absolute value of the acceleration parameter $\ddot a/a$ for the previous cases as a function of time. Knowing that the evolution starts with a period of inflation, time intervals of decelerated expansion can be identified from this plot. They correspond to the steep dips in $\ddot a/a$, where two sign changes of the acceleration parameter occur. There is no interruption of accelerated expansion in the second case. This is the reason for the field to be trapped at $\phi\approx 0$ entailing a sustained accelerated expansion.
%
%

\section{Scalar and tensor perturbations}\label{s:Fluctuations}

\begin{figure}
	\centering
	\includegraphics[scale=0.65]{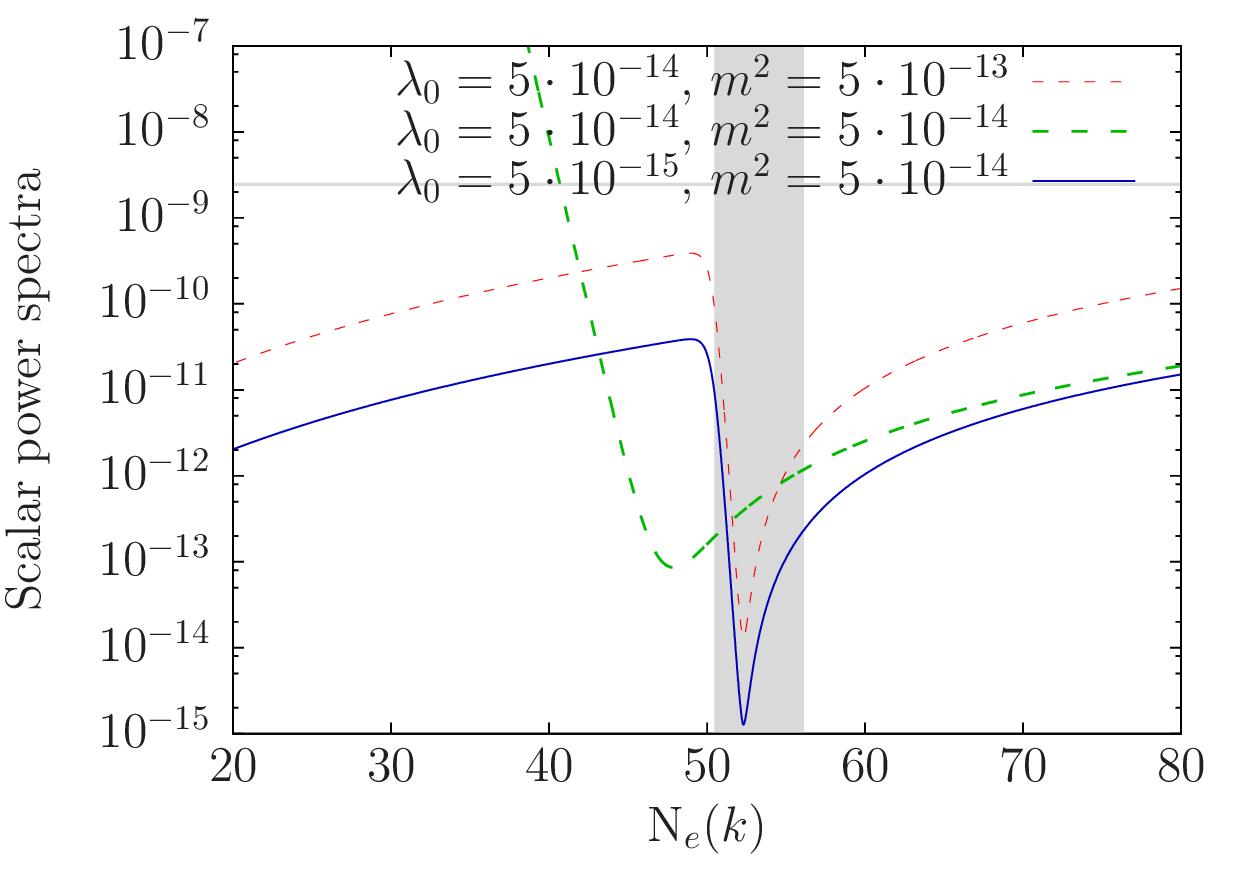} 
	\caption{Resulting scalar power spectra $\mathcal P_\mathcal R (k)$ in the same three cases as in Figs.~\ref{fig:t410b_QuadDilFields_6-7-10_bw} and \ref{fig:t412_QuadDiladdot_6-7-10}. As can be seen from the cases one and three, shifting the potential by a constant factor leads to the same shift in the spectrum.} 
\label{fig:t420_QuadDilSSpectra_6-7-10}	
\end{figure}

\begin{figure}
	\centering
	\includegraphics[scale=0.65]{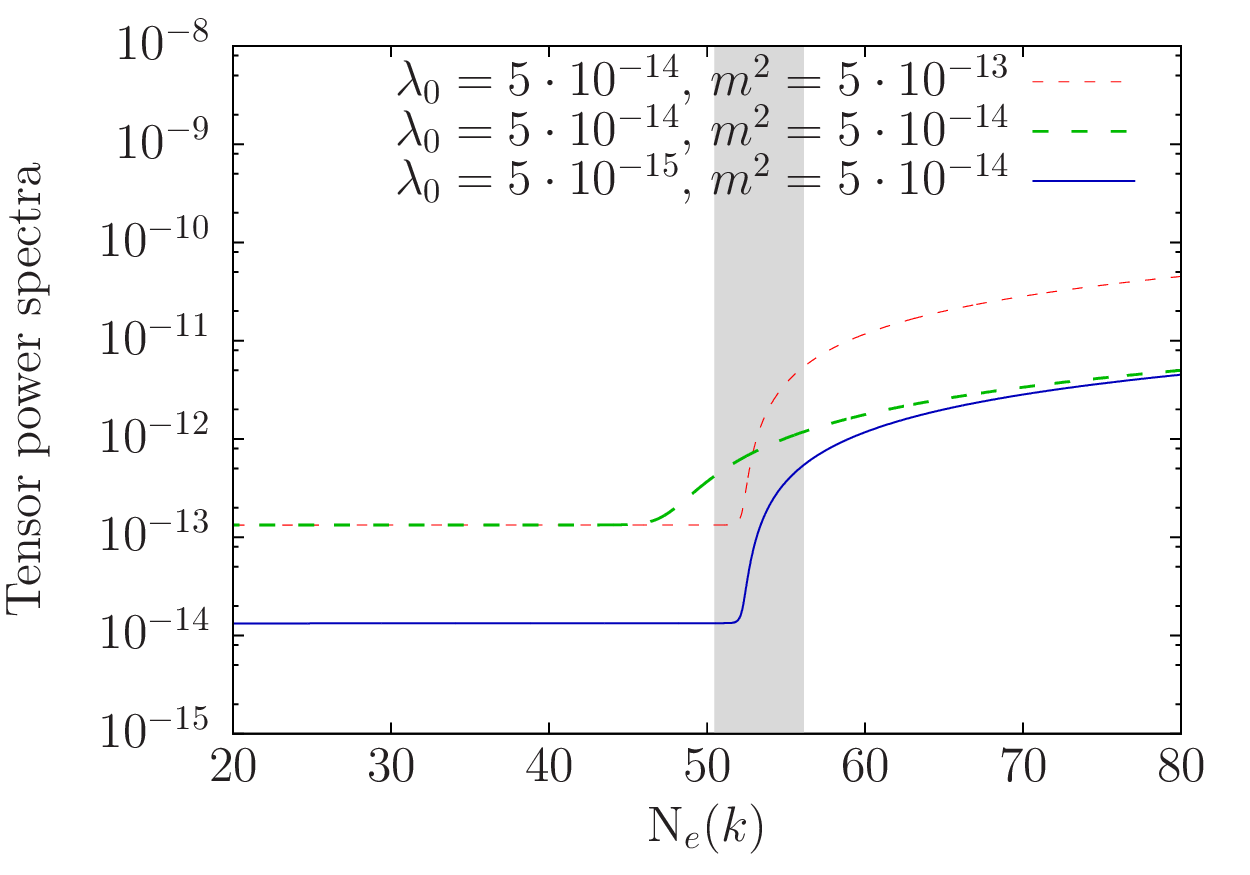} 
	\caption{Tensor power spectra $\mathcal P_T (k)$ in the same three cases as in Figs.~\ref{fig:t410b_QuadDilFields_6-7-10_bw} to \ref{fig:t420_QuadDilSSpectra_6-7-10}. As can be seen from the cases one and three, shifting the potential by a constant factor leads to the same shift in the spectra. The line corresponding to the second parameter set interpolates between the two other cases because the respective potentials are (almost) the same at the relevant instances of time.}
\label{fig:t421_QuadDilTSpectra_6-7-10}	
\end{figure}

The spectra produced in this scenario are plotted against $N_e$ in 
Figs.~\ref{fig:t420_QuadDilSSpectra_6-7-10} and~\ref{fig:t421_QuadDilTSpectra_6-7-10}. Here $N_e$ is the number of e-foldings between horizon exit of the mode $k=2\pi/\lambda$ and the end of inflation. It is defined in terms of the corresponding scale parameters as
\begin{equation}
	N_e = \text{ln} \frac{a_\text{end}}{a(k)}
\end{equation}
and can be thought of as a function of the wavenumber $k$. It can be also expressed as a function of the field value, which has been done using the slow-roll approximations
\begin{equation}
	\frac{\dot\phi^2}{2V(\phi)}\ll1 \quad\text{and}\quad \left|\frac{\ddot\phi}{3H\dot\phi}\right| \ll 1
\end{equation}
with the result
\begin{equation}
	N_e(\phi) =  \text{ln} \frac{a_\text{end}}{a(\phi)} = \frac{8\pi}{m^2_\text{Pl}} \int_{\phi_\text{end}}^\phi \frac{V}{V'}d\phi,
\label{eq:NeOfPhi}
\end{equation}
where, as before, the label ``end'' refers to the end of inflation. For the parameter sets one and three the corresponding potentials are proportional to each other, which is the reason why the function $N_e(\phi)$ is identical in both cases as long as the slow-roll assumption is valid, see Eq.~\eqref{eq:NeOfPhi}. So the transition from the first to the second period of inflation takes place at the same value of $N_e$, which is obtained to be $N_e\approx 50$. Even for larger $N_e$ this proportionality holds, which suggests that the short time interval of slow-roll breaking is not important here.
Then it follows from the formula for the scalar power spectrum in slow-roll approximation,
\begin{equation}
	\mathcal P_\mathcal R (k) = \left( \frac{H^2}{2\pi \dot\phi_c}\right)^2_{t_k} = \frac{128\pi}{3} \frac{V^{3}}{m^6_\text{Pl}V'^2},
\end{equation}
that also the two scalar power spectra should be proportional to each other. The same equation explains the diverging behavior of the scalar power spectrum in the second case of Fig.~\ref{fig:t420_QuadDilSSpectra_6-7-10}, where $\dot\phi$ vanishes at the transition to the second inflation. The curve referring to the second case is given an arbitrary offset in $N_e(k)$-direction. Because there is no end of inflation, it cannot be uniquely fixed.

Figure~\ref{fig:t421_QuadDilTSpectra_6-7-10} shows the corresponding tensor power spectra which in the slow-roll approximation are given by
\begin{equation}
	\mathcal P_T = \frac{128}{3} \frac{V}{m_\text{Pl}^4}.
\label{eq:TensorPSSlowRoll}
\end{equation}
Being proportional to the Hubble parameter at horizon exit of the mode in question, the curves just illustrate the loss of (mostly) potential energy of the field.   
%
%
The scalar and the tensor power spectra arising from the potential $V_4$ show similar properties and are not displayed. Both of the potentials $V_2$ and $V_4$ allow for upward and sideward shifts of the power spectra by choosing corresponding parameter values.

\begin{figure}
	\centering
	\includegraphics[scale=0.55]{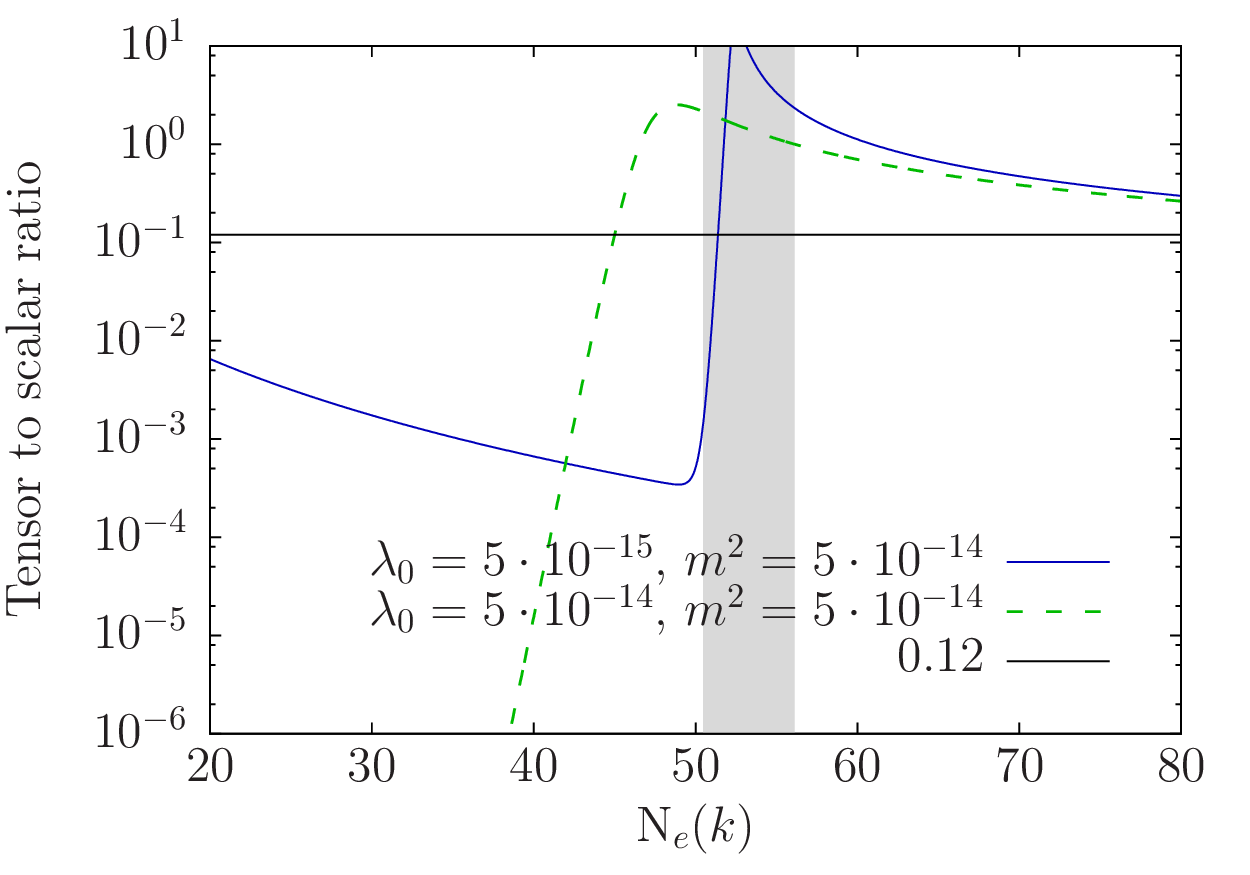} 
	\caption{The tensor-to-scalar ratio strongly drops down at the transition to small-field inflation. If $\dot\phi$ becomes very small, also $r$ vanishes at corresponding values of $N_e(k)$ (left panel, green curve).}
\label{fig:t424_t425_QuadDil_r_6-7-10}	
\end{figure}

\begin{figure}
	\centering
	\includegraphics[scale=0.55]{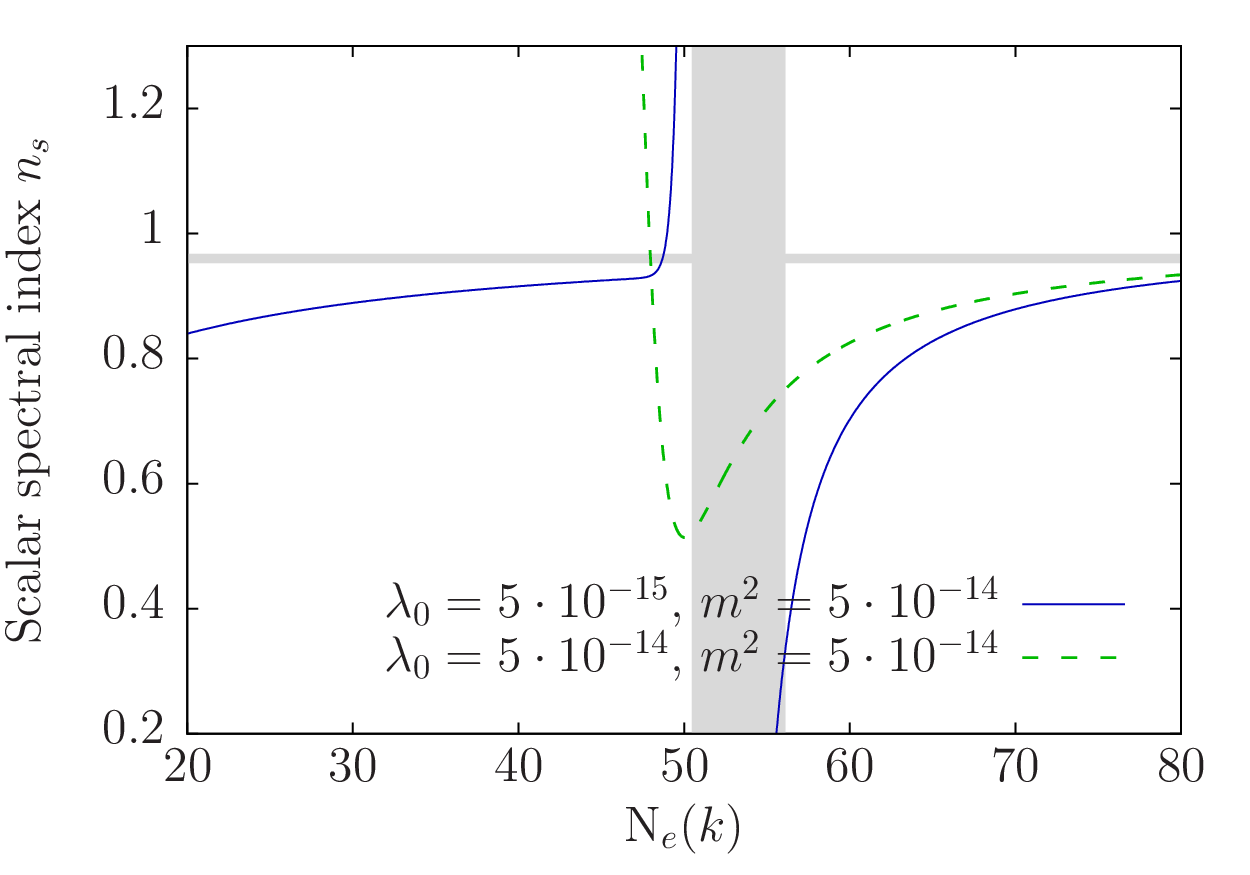} 
	\caption{Evolution of the scalar spectral index for the previously discussed cases. The transition region between the two periods of inflation should be far outside the sensitivity range of the Planck measurements. The scalar spectrum within the third case in Fig.~\ref{fig:t420_QuadDilSSpectra_6-7-10} was just shifted by a factor with respect to the first one. As expected, the logarithmic derivative of these functions is the same, and therefore omitted in this figure.}
\label{fig:t426_QuadDil_n_s_6-7-10}	
\end{figure}

The parameters for Figs.~\ref{fig:t410b_QuadDilFields_6-7-10_bw} to~\ref{fig:t426_QuadDil_n_s_6-7-10} are chosen such that the wavenumber $q_0=0.002 \text{Mpc}^{-1}$ today corresponds to a mode that exits the horizon around the transition between the two inflationary periods. 
The correspondence to $q_0$ cannot be narrowed down to a single value of $N_e$ or $k$ because the physics of the inflaton decay is not settled. 
Altogether, CMB measurements cover an $N_e$--range of roughly ten \cite{Bringmann12}. 
For these scales the primordial fluctuations have been obtained to high accuracy. 
So, for the presented scenarios to be viable, there should be a range of at least ten e-foldings in which the computational results match with observation. 
This range should overlap with the grey band in  Figs.~\ref{fig:t420_QuadDilSSpectra_6-7-10} to~\ref{fig:t426_QuadDil_n_s_6-7-10}.
The observational result \footnote
{While observation does not yield exactly the same value for the whole interval, compared to the large gradients of the calculated curves it can be regarded as constant.}
is represented by the narrow horizontal stripe at $2.4\cdot 10^{-9}$. While it is possible to tune the parameters such that the results match for a fixed wavelength, the strong variation of $\mathcal P_\mathcal R (k)$ on the scale of a few $N_e(k)$ renders it impossible to fit the curves to the measurements over the whole interval. This leads to the conclusion that the transition between large-field and small-field inflation should occur when the Hubble parameter is either much smaller or much larger than the wavenumber $q_0$ today. 

The tensor-to-scalar ratio in slow-roll approximation is
\begin{equation}
	r = \frac{\mathcal P_T }{\mathcal P_\mathcal R} = \frac{m_\text{Pl}^2}{\pi}\left( \frac{V'}{V} \right)^2.
\label{eq:rSlowRoll}
\end{equation}
Results for $r$ within the potential $V_2$ are depicted in Fig.~\ref{fig:t424_t425_QuadDil_r_6-7-10}. During the transition to small-field inflation, $r$ drops by several orders of magnitude. This reflects the drop of the field velocity $\dot\phi$. It occurs because the Hubble drag is the only force acting on the field in an almost flat region of the potential. The cases one and three in Fig.~\ref{fig:t420_QuadDilSSpectra_6-7-10} yield the same (solid) curve here. The dashed curve drops to very small numbers when $\phi$ is practically fixed at a constant value. Also for $r$, the maximally allowed value of $r=0.12$, shows that compatibility is only possible when $\phi=0$ is reached outside the range observable in the CMB. 
%
%

\begin{figure}
	\centering
	\includegraphics[scale=.7]{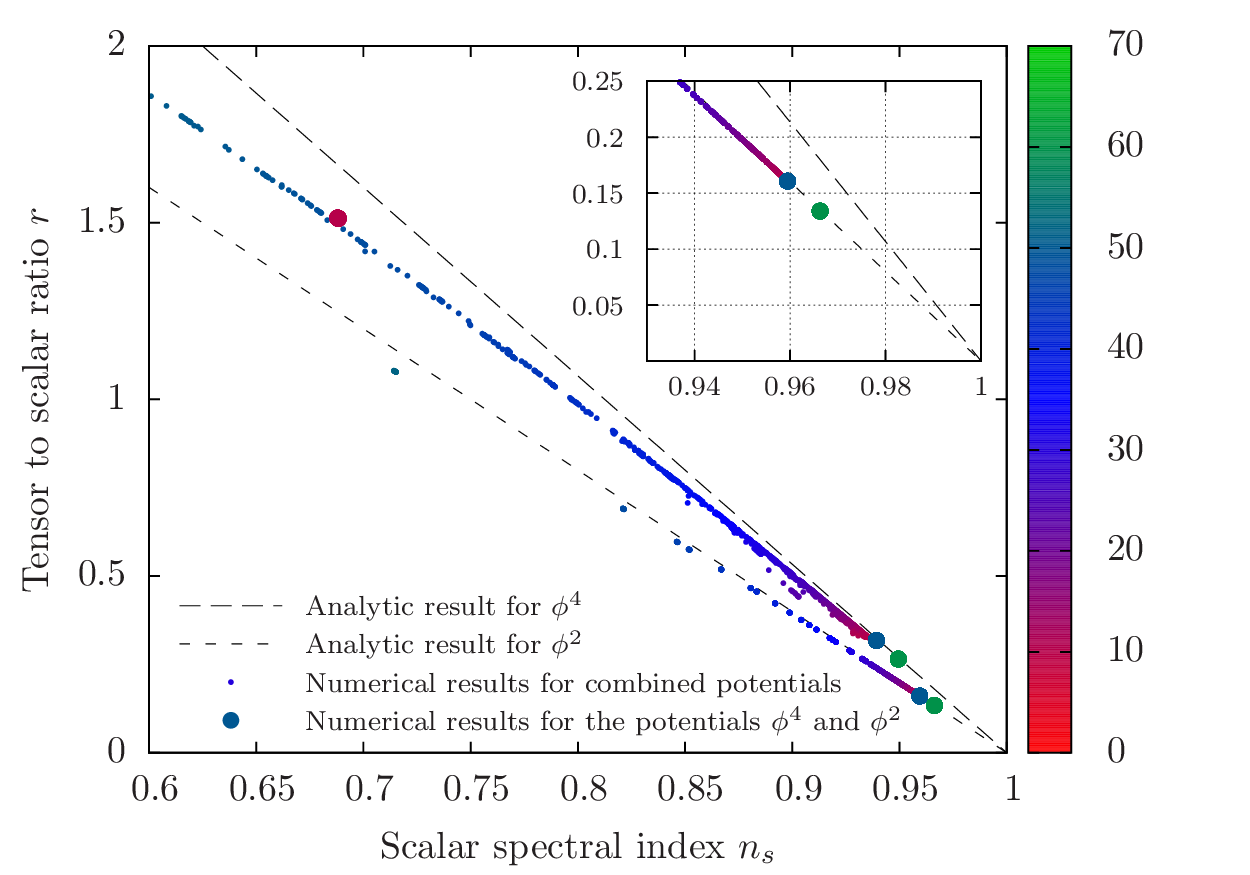} 
	\caption{This figure shows the result of different calculations of the tensor-to-scalar ratio $r$ and the scalar spectral index $n_s$. These quantities are computed for modes leaving the horizon a fixed number of e-foldings before inflation ends. The two dashed lines follow the analytic formulae \eqref{eq:r_of_ns} for slow-roll inflation within the potentials $m^2\phi^2/2$ and $\lambda \phi^4/4$, respectively. The large dots mark the results of numerical calculations within the same two potentials. The number of e-foldings $N_e$ before the end of inflation are $60$, $50$, and $10$, respectively. Along the dashed lines (and more and more departing from them) the small dots represent results from the combined potentials $V_2$ and $V_4$ described in the main text. The color code shows the number of e-foldings of the second inflation, whereas the total $N_e$ is $60$ for each of the small dots. All results were obtained within slow-roll approximation.}
\label{fig:t500_Ne2ColorCodeAllModels}	
\end{figure}

\begin{figure}
	\centering
	\includegraphics[scale=.7]{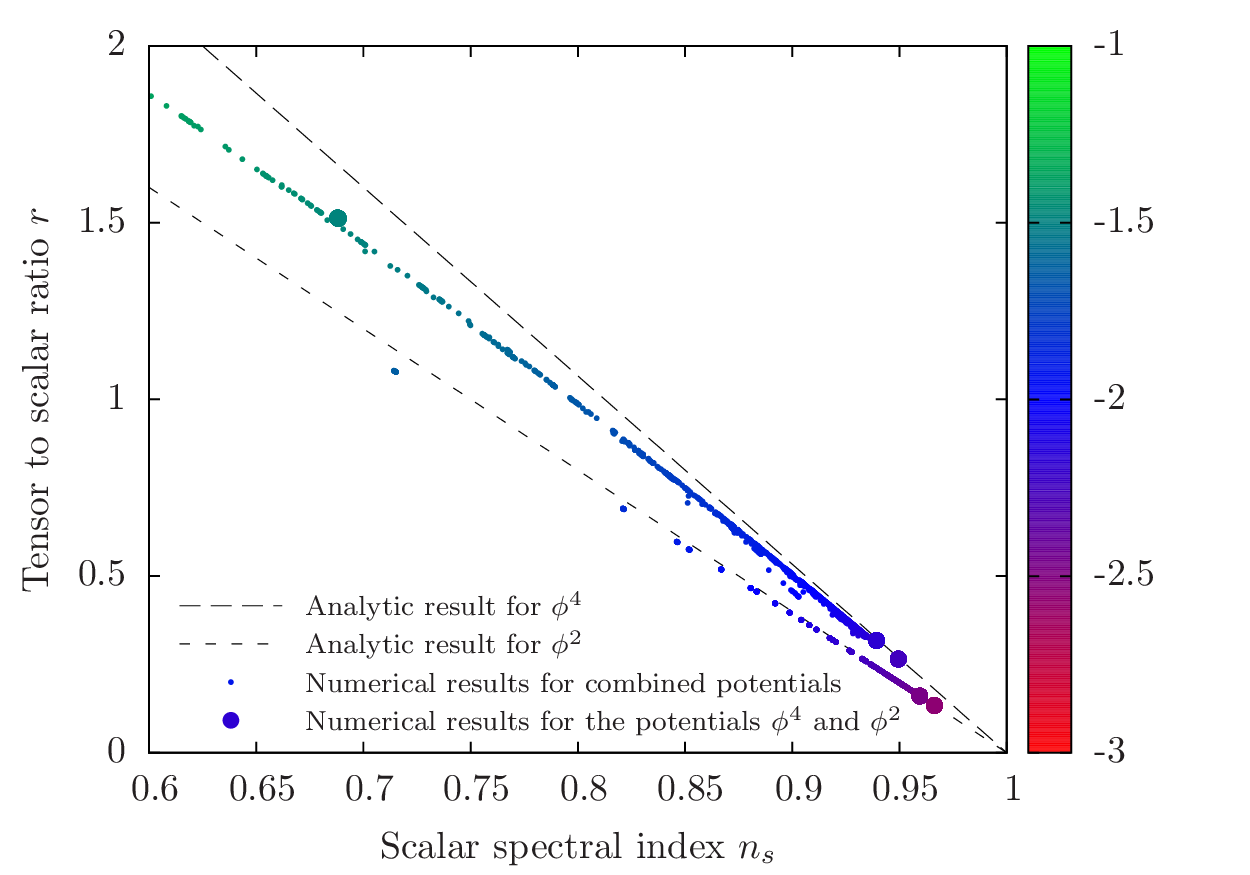}
	\caption{The same as in Fig.~\ref{fig:t500_Ne2ColorCodeAllModels} but with the colors representing the decimal logarithm of the slow-roll condition $\dot\phi^2/2V(\phi)$. This figure shows that the deviation of the points from the dashed lines can be explained from the lack of slow roll.}
\label{fig:t501_sr2ColorCodeAllModels}	
\end{figure}

\begin{figure}
	\centering
	\includegraphics[scale=.7]{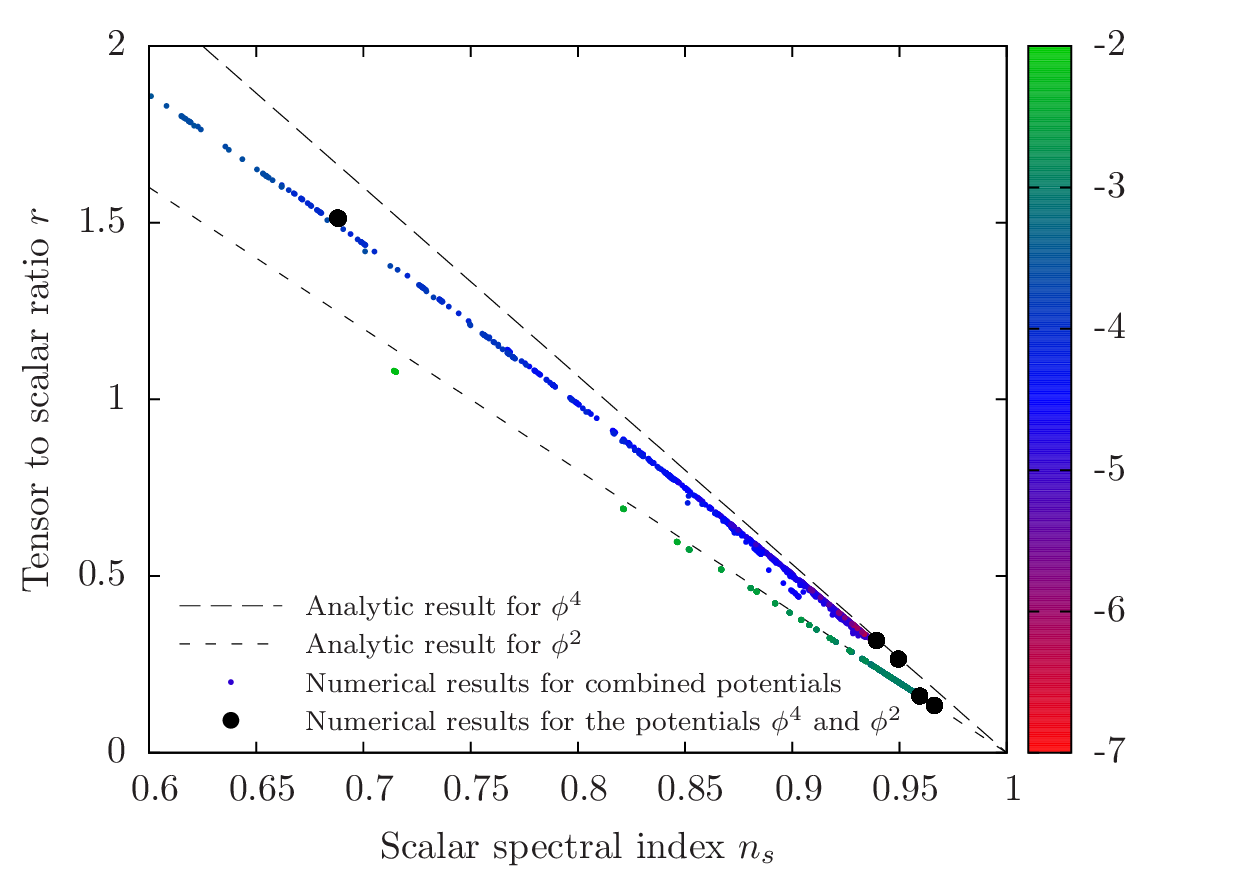} 
	\caption{The same as in Figs.~\ref{fig:t500_Ne2ColorCodeAllModels} and~\ref{fig:t501_sr2ColorCodeAllModels}  but with the colors representing the decimal logarithm of the quantities $V_0/m^2\phi^2$ and $4V_0/\lambda_1\phi^4$ for the combined potentials with $m^2\phi^2/2$ and with $\lambda_1\phi^4/4$, respectively. In combination with Fig.~\ref{fig:t500_Ne2ColorCodeAllModels} this figure suggests that the deviation of the numerical calculation from the analytic result stems from the violation of slow roll rather than from the offset $V_0$ in the potential $V(\phi)$.}
\label{fig:t502_offsetColorCodeAllModels}	
\end{figure}

The same is true for the allowed range of the scalar spectral index $n_s$, which can be seen in Fig.~\ref{fig:t426_QuadDil_n_s_6-7-10}. During the transition it takes on values far away from the ones being measured. It is clear from the definition of $n_s$,
\begin{equation}
	n_s = \frac{d \ln  \mathcal P_\mathcal R(k)}{d \ln k} + 1,
\label{eq:Def_ns_nT}
\end{equation}
that multiplying the scalar spectrum with a constant factor does not alter the value of the spectral index. So parameter set three yields the same result as set one and is not displayed.
Within the slow-roll approximation, $n_s$ can be expressed by the slow-roll parameters 
\begin{equation}
	\epsilon = \frac{m_\text{Pl}^2}{16\pi} \left( \frac{V'}{V} \right)^2 
\quad \text{and} \quad 
	\eta = \frac{m_\text{Pl}^2}{8\pi} \left( \frac{V''}{V} \right)
\end{equation}
as
\begin{equation}
	n_s = 2\eta - 6\epsilon.
\label{eq:ns_eta_epsilon}
\end{equation}
This is used in Figures~\ref{fig:t500_Ne2ColorCodeAllModels} to~\ref{fig:t502_offsetColorCodeAllModels} to summarize the results for the tensor-to-scalar ratio and for the scalar spectral index.
%
%
As noted in the keys of these figures, the black dashed lines sketch the dependence of $r$ on $n_s$ as obtained from an analytical estimation for the two potentials $m^2\phi^2 /2$ and $\lambda \phi^4/4$. The result varies along these lines when different values of $N_e$ are chosen. Including the possibility of an offset $V_0$, the slow-roll equations~\eqref{eq:rSlowRoll} and~\eqref{eq:ns_eta_epsilon} for the scalar spectral index combine to 
\begin{equation}
	r_2 = 4 \frac{n_s - 1}{V_0/m^2\phi^2-1} \quad \text{and}\quad 	r_4 =  \frac{16}{3}\frac{n_s - 1}{4V_0/{\lambda}\phi^4-1},
\label{eq:r_of_ns}
\end{equation}
where $r_2$ and $r_4$ are the tensor-to-scalar ratio in the squared and the quartic case, respectively. The black dashed lines give the results of Eq.~\eqref{eq:r_of_ns} with $V_0$ set to zero.
Numerical calculations within the same cases are represented by the five large dots which are obtained at three different stages of inflation. These are labelled by the number $N_e$ which is large ($N_e=50$ and 60) for early production well within the slow-roll regime and small ($N_e=10$) for late production, i.e. for fluctuations in modes crossing the horizon shortly before the end of inflation. Then the slow-roll approximation is expected to lose its validity. So, the deviation of the large dot in dark red from the long-dashed black line is no surprise since also the dots are calculated assuming slow roll. The position of the large dots is independent from the coupling when values $\lambda \in [10^{-15},10^{-11}]$ and $m^2 \in [10^{-15},10^{-11}]$ are chosen. The small dots represent results computed within the combined potentials $V_2$ and $V_4$. They arrange along curves starting on the corresponding straight line (for the monomic quadratic and quartic potential, respectively) and more and more deviate when evolving to larger $r$ and smaller $n_s$. Since this deviation also occurs for the monomic potential, it is attributed to the breaking of slow roll. Corrections due to $V_0$ seem to be small in comparison. They tend to enlarge the value of $r$ at given $n_s$.
 
%
%

The color of the small dots encodes the number of e-foldings of the second inflation, which varies according to different parameter choices. It is clearly visible that this number $N_e^{(2)}$ grows monotonically when following the dots from the lower right to the upper left. This suggests the following interpretation: The fluctuations being measured within the present scenario at $N_e = 60$ resemble those that would be produced in the quartic and quadratic potentials at later times. The main effect of the second period of inflation is a shift of the remaining e-foldings connected to the fluctuations.

The small figure in Fig.~\ref{fig:t500_Ne2ColorCodeAllModels} is a detail of the lower right region of the large one. This is the interesting range as CMB measurements are concerned. The values allowed by Planck lie below $0.15$ for $r$ and between $0.94$ and $0.98$ for $n_s$. So one can see that the second inflation drives the computed values of $r$ and $n_s$ further away from the ones measured in the CMB. The more effective it is in terms of e-foldings the more severe it renders the discrepancy.   
%
%

Figs.~\ref{fig:t501_sr2ColorCodeAllModels} and~\ref{fig:t502_offsetColorCodeAllModels} show the same results as Fig.~\ref{fig:t500_Ne2ColorCodeAllModels} but with a different choice of the color code. This is used to illustrate the reason why the lines formed by the small dots deviate from the analytic estimate (dashed lines). In Fig.~\ref{fig:t501_sr2ColorCodeAllModels} the colors reflect the value of the slow-roll measure $\log_{10} (\dot\phi^2 / 2V(\phi))$. The largest deviations are observed where the kinetic energy amounts to a few percent of the potential and the results agree better for a lower proportion. 

The values of $V_0/m^2\phi^2$ and $4V_0/\lambda_1 \phi^4$ are encoded in the colors of the dots in Fig.~\ref{fig:t502_offsetColorCodeAllModels}. These quantities are proportional to the ratio of the offset to the field dependent part of the potential, and they are computed at the time the mode of interest leaves the horizon. They also occur in Eq.~\eqref{eq:r_of_ns}. As stated above, they are expected to give corrections enlarging $r$ for given $n_s$. So it can be concluded that their influence on the result is less important than that of missing slow roll. This is confirmed by the fact that the deviations are considerably smaller for the massive field, where $V/m^2\phi^2$ is largest and slow-roll violation is quite small. For the pure squared and quartic potentials the quantity reflected by the color code is zero and they are plotted in black. 

The parameter sets that yield the small dots are chosen in the following way: First a sweep over the parameter range $v \in [0.01,1]\cdot m_\text{Pl}$,  $\lambda_0 \in [10^{-15}, 10^{-12}]$, $\lambda_1 \in [10^{-15}, 10^{-12}]$ is done. The parameter $\lambda_2$ is restricted to $\lambda_2 = 0$ and the interval $\lambda_2 \in [10^{-15}, 10^{-14}]$.  Stronger couplings are not computed with because they lead to larger values of $V_0$ which stop the evolution of $\phi$ before reaching the minimum. 

For each parameter set the e-foldings of the first and the second inflation, $N_e^{(1)}$ and $N_e^{(2)}$, are determined. The interesting cases are those with enough $N_e^{(2)}$, such that the expected spectra are discernible from standard large field inflation. On the other hand the value should not exceed $N_e^{(2)} = 60$ because then only the second, small-field inflation is visible today. 
As exemplified in Figs.~\ref{fig:t420_QuadDilSSpectra_6-7-10} to~\ref{fig:t426_QuadDil_n_s_6-7-10} the spectra produced near the transition between the two inflationary periods deviate strongly from the CMB measurements which set tight bounds on the magnitude of fluctuations. These arguments lead to the approach that parameter sets entailing $N_e^{(2)}\in[10,60]$ are selected and the fluctuations are calculated for modes leaving the horizon at $N_e=60$ e-foldings before the end of the second inflation.
%
%

\section{Fluctuations produced after inflation: Preheating}\label{s:Preheating}

At the end of inflation, the universe is assumed to be cold and empty, with the exception of the inflaton field. 
The inflaton pervades space almost homogeneously before it decays into its excitations and other particles. 
The first, rapid decay is called preheating and it is characterized by non-perturbative effects. 
It is crucial for accomplishing thermal equilibrium at the temperature, where Big Bang nucleosynthesis takes place. 
This is decisive for a scenario to be viable.
This work is only concerned with preheating while skipping the following period of reheating which is dominated by perturbative effects.

\begin{figure}
	\centering
	\includegraphics[scale=0.65]{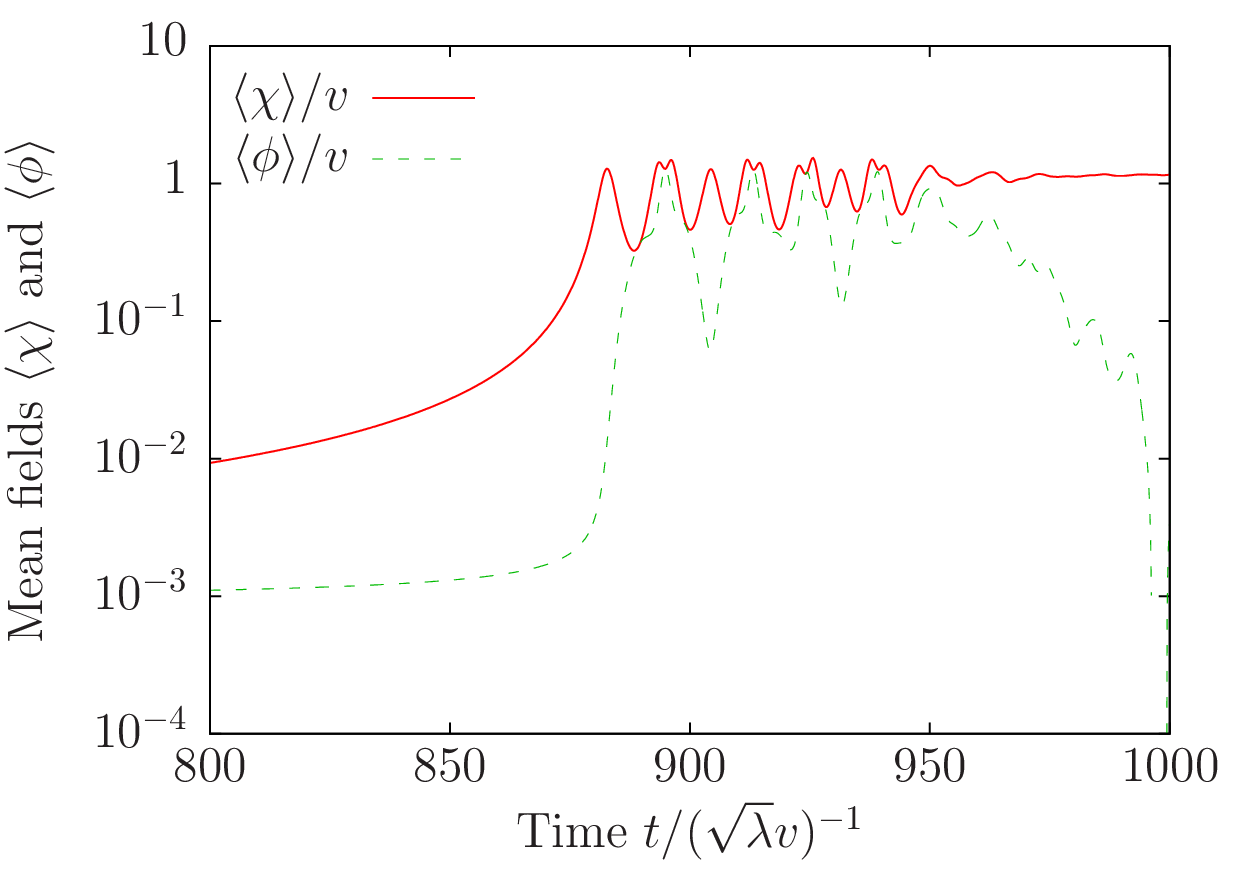}
	\caption{Evolution of the means $\langle\chi\rangle$ and $\langle\phi\rangle$ after the end of inflation. In this calculation both fields start at $10^{-3}v$ with zero initial velocity. After a few oscillations the zero mode of $\chi$ is at rest in the potential minimum. The mean of $\phi$ can also switch between different values of minimal energy.
	Here and for the following Figures~\ref{fig:t641_PhiLogVars} and~\ref{fig:t630_PhiLogSpectra}, the couplings are $\lambda_0=\lambda_1=\lambda_2=10^{-13}$ and the energy scale is $v=10^{-3}m_\text{Pl}$. Field fluctuations are included.}
\label{fig:t640_PhiLogMeans}
\end{figure}

\begin{figure}
	\centering
	\includegraphics[scale=0.56]{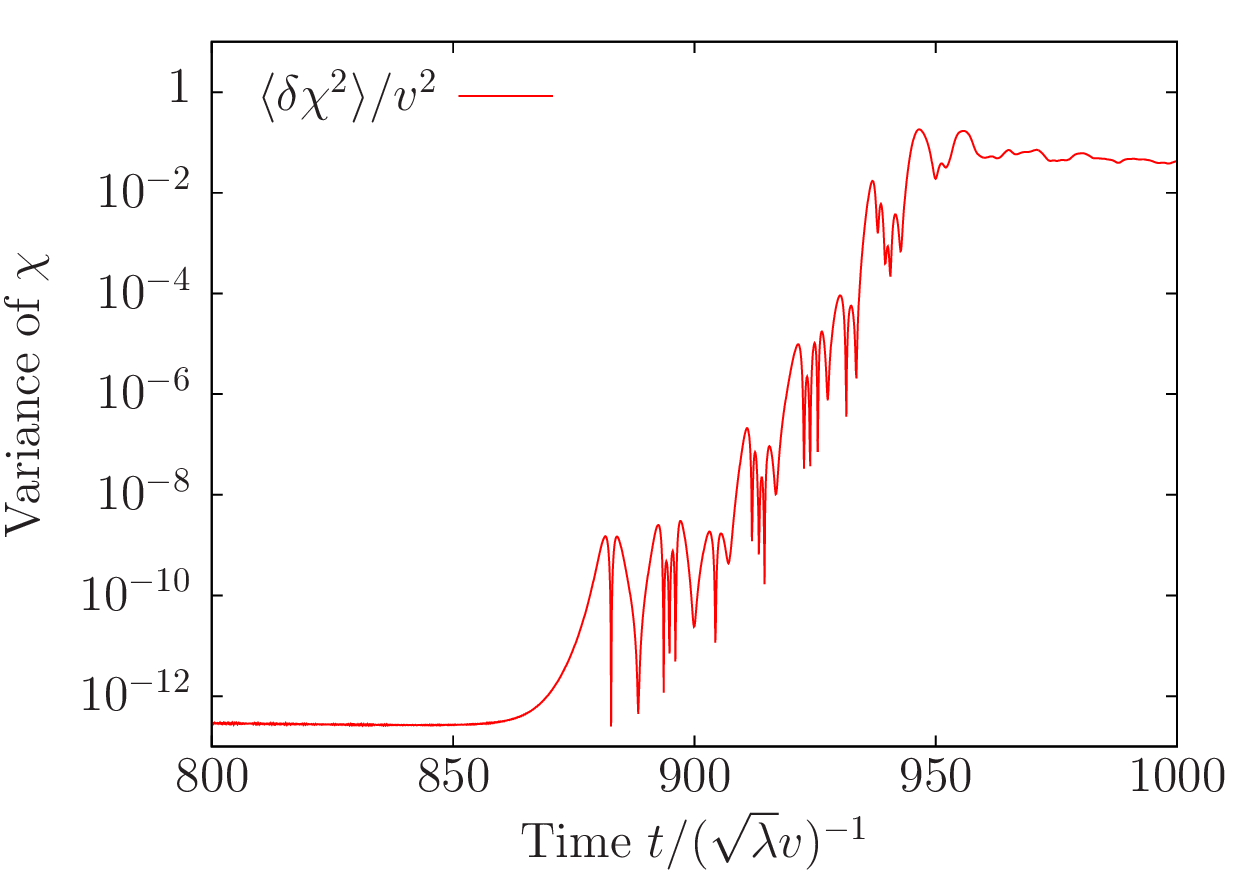}
	\includegraphics[scale=0.56]{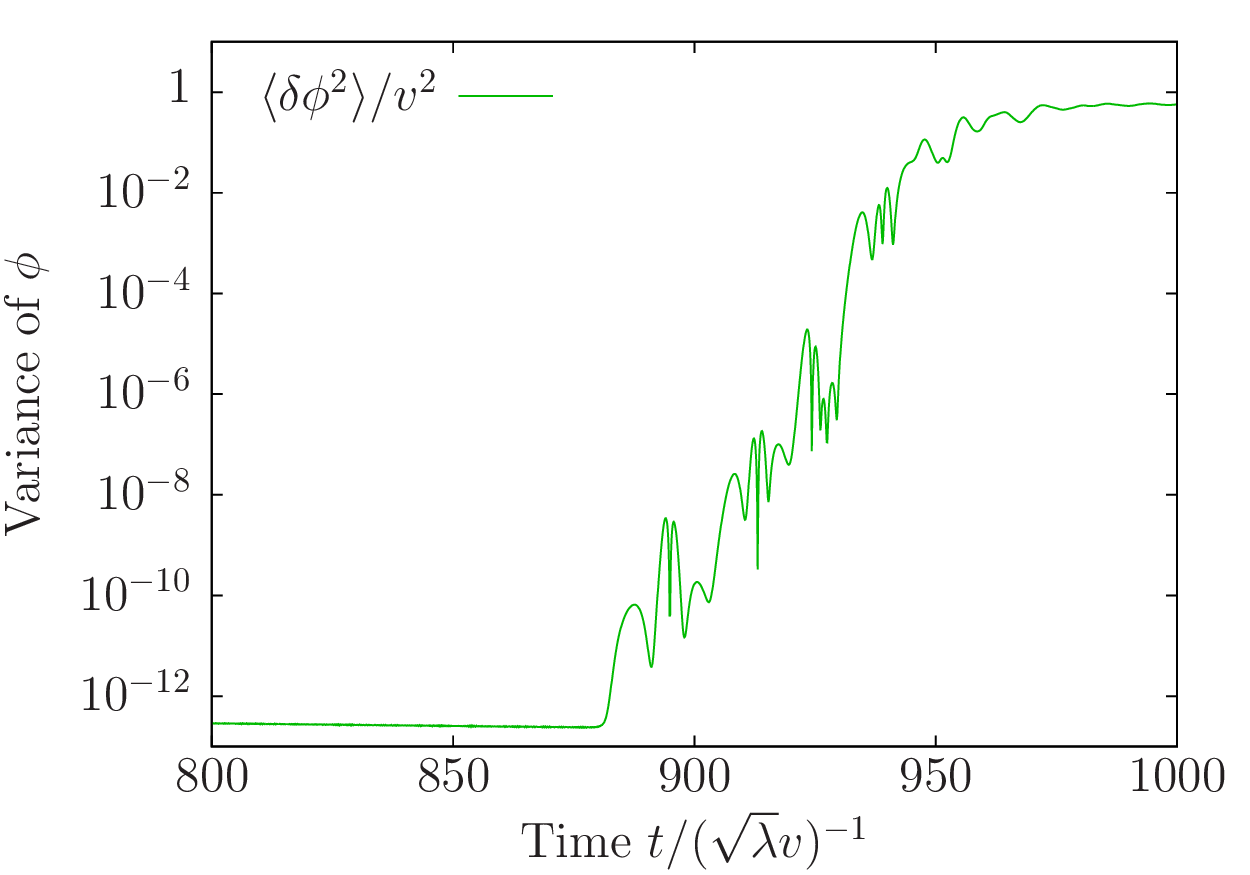}
	\caption{Time dependence of the field variances in the same case as in Figs.~\ref{fig:t640_PhiLogMeans} and~\ref{fig:t630_PhiLogSpectra}. The variance of $\chi$ follows the growth of the zero mode. Inhomogeneities of $\phi$ evolve with a short delay, which is also observed for the mean field.}
\label{fig:t641_PhiLogVars}
\end{figure}

\begin{figure}
	\centering
	\includegraphics[scale=0.65]{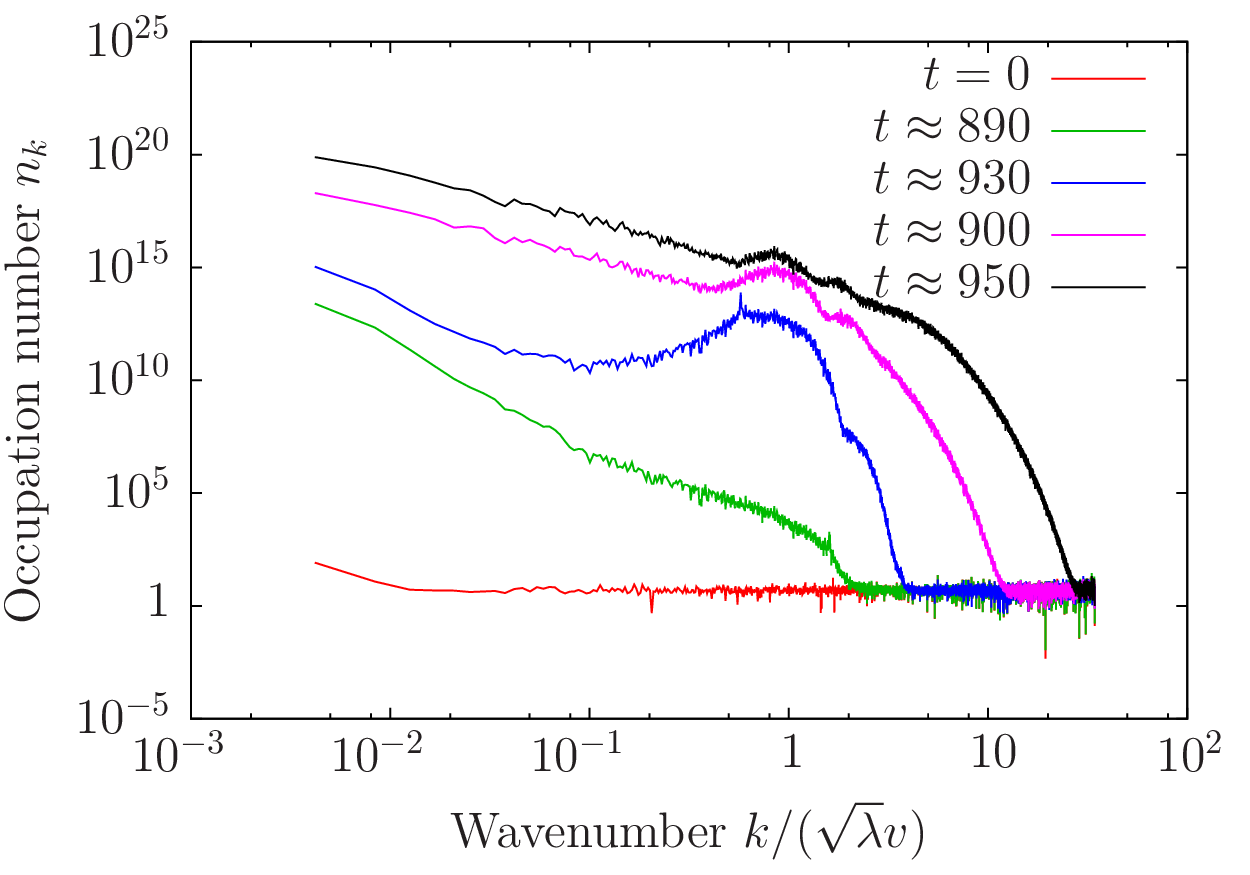}
	\caption{Spectra of $\chi$ particles from the same calculation as discussed in Figs.~\ref{fig:t640_PhiLogMeans} and~\ref{fig:t641_PhiLogVars}.  The spectra of the field $\phi$ are very similar and are therefore omitted. As in the case without coupling to a second scalar, a temporary maximum forms at slightly lower momenta than $\sqrt\lambda v$, which is the effective mass squared of $\chi$ in the minimum.}
\label{fig:t630_PhiLogSpectra}
\end{figure}

In Ref.~\cite{DesrocheFKL05} preheating in the potential
\begin{equation}
	V(\phi) =  \frac{1}{16}\lambda_0 v^4 + \frac{1}{4} \lambda_0 \chi^4 \left(\ln \left|\frac{\chi}{v}\right|-\frac{1}{4}\right),
\label{eq:NewInfl}
\end{equation}
is discussed. The numerical calculation is done with the C++ program LATTICEEASY by G. Felder and I. Tkachev \cite{FelderT08}. For the present case of the potentials Eqs.~\eqref{eq:SimplPotPhi4}, and~\eqref{eq:SimplPotPhi2} this program has also been used, and similar results have been obtained:							

The second inflationary period ends somewhere not too far from the origin in field space, the fields being on their way to one of the minima of the potential. This situation is similar to preheating in the potential Eq.~\eqref{eq:NewInfl} with an additional coupling to a second field. The evolution of the homogeneous modes and the particle spectra of both fields resembles the results in Ref.~\cite{DesrocheFKL05}.
Figs.~\ref{fig:t640_PhiLogMeans} to~\ref{fig:t630_PhiLogSpectra} depict the results of a lattice calculation using this potential. 

The calculation is done within the parameter set 
$\lambda\equiv\lambda_0=\lambda_1=\lambda_2=10^{-13}$ and $v=10^{-3}m_\text{Pl}$. Following Ref.~\cite{DesrocheFKL05} the computation has been done in one spatial dimension.
%
%
The initial values of the fields are both set to $10^{-3}v$. Presuming slow roll, the initial field velocities are set to zero. From Fig.~\ref{fig:t640_PhiLogMeans} it is seen that the mean values of the fields start to grow and later oscillate around one of the minima
\begin{equation}
	\chi_\text{min} = \pm v \exp\left(\frac{\lambda_2^2}{4\lambda_0\lambda_1}\right),  \quad 
	\phi_\text{min} = \pm \sqrt{\frac{\lambda_2}{2\lambda_1}}\chi_\text{min}.
\label{eq:PotMinimum}
\end{equation}
 %
%
Comparison with Fig.~\ref{fig:t641_PhiLogVars} shows that the oscillation around these values ends when the variances have grown to almost unity. 
In Ref.~\cite{AntuschNO15} it is shown that for two spatial dimensions there are regions in space where the field initially oscillates between the two vacuum values.

Fig.~\ref{fig:t630_PhiLogSpectra} contains the spectra of inhomogeneities of the field $\chi$ at some moments of time. The time unit is the time scale of oscillations around the minimum, $(\sqrt\lambda v)^{-1}$. The noise of the spectra is reduced by averaging over some sets of different initial vacuum fluctuations. 

The shape of the spectra differs only in details from the ones shown in Ref.~\cite{DesrocheFKL05}. The characteristic momentum scale of the transient maximum of the spectrum shows the same dependence on the parameters of the potential. In both cases the scattering of particles out of the maximum appears as a small amplification of the modes with the approximately double momentum. 
Scattering processes drive the front of the spectrum to higher and higher momenta, finally leading to thermal equilibrium.
For the field $\phi$ no spectra are displayed because they show no significant differences to the corresponding spectra of $\chi$. As expected, by varying the couplings $\lambda_i$ the location of the resonance scale can be moved to other values.

%
%
The calculation of preheating in the potential Eq.~\eqref{eq:FullPotPhiChi} has also been done for higher energy scale $v=10^{-1}m_\text{Pl}$. As in the case without additional field, no efficient particle production could be observed. The variances did not grow from very small values for various combinations of the couplings.

\section{Conclusion}\label{s:Conclusion}

This paper presents results on the behavior of fields and particles during and after two subsequent stages of cosmological inflation. The calculations make use of the slow-roll approximation which is justified in two regions of the potentials Eq.~\eqref{eq:SimplPotPhi2} and Eq.~\eqref{eq:SimplPotPhi4} for a large parameter space. These potentials are simplified forms of one that has been used to describe strongly interacting matter with effective fields.  

It is based on the linear sigma model, which incorporates chiral symmetry, and an additional scalar dilaton field has been introduced to imitate the QCD-behavior under scale transformations.
In Ref.~\cite{BoeckelS10}, a dilaton-extended linear sigma model has already been used for an inflationary scenario during the cosmological QCD phase transition. In the work at hand, the consequences of a similar potential during primordial inflation well before the QCD phase transition has been examined. 

In this potential, the early stages of inflation are very similar to the corresponding ones in the quartic and quadratic potentials of large-field inflation. Also the primordial spectrum of fluctuations that originates from this epoch does not differ significantly from the spectra produced by these inflationary models. However, as in other models of hybrid inflation, at the end of this period the fields have not reached the minimum of the potential. This occurs only after an additional waterfall stage that can bring about a considerable number of e-foldings: In Ref.~\cite{Clesse11} such a scenario has been discussed for large amounts of inflation during the waterfall period. Then all observational inflation occurs in a small-field setting. 

By contrast, in Ref.~\cite{ClesseR14} the second inflation is assumed to be short or absent, such that no additional e-foldings of a second inflation have to be taken into account. 
Then, the first possibility is a result equivalent to chaotic inflation: 
This may happen when the offset of the potential has negligible effect, as it has been seen for the calculations in this paper. It corresponds to the simplified picture of an inflaton rolling down a potential $V_0+\lambda\phi^n/n$, where the potential is immediately set to zero when $\phi = 0$ is reached.

A second possibility is the following: The slow roll of $\phi$ becomes unstable much earlier. Then inflation stops at a position in the potential $\phi_\text{end}$ for which  $N_e(\phi_\text{end})>1$ when computed in the respective monomic potential. The resulting spectrum resembles that from $\lambda \phi^n/n$ but its momentum scale today is shifted. 

The spectra resulting from the calculations of this work are shifted, too, but towards the opposite direction: The scenario with two inflationary epochs prolongs inflation after its natural end. Instead of being interrupted early, inflation continues (after a short pause) and thus shifts the fluctuations to larger wavelengths than expected. So, the fluctuations at some physical $q_0$ today are similar to those which are expected from monomic inflation at some smaller momentum scale. They will enter the horizon at some time in the future. In other words: The fluctuations obtained from CMB measurements and being linked to a certain number of e-foldings subsequent to their horizon exit, exhibit features, which are expected from fluctuations connected to some significantly smaller value of $N_e$, when a monomic potential is assumed to drive inflation. The case with negligible second inflation ($N_e^{(2)}\ll10$) gives spectra indistinguishable from the $\lambda \phi^n/n$ standard scenario and is therefore omitted from further discussion.
It has been justified that the discussion concentrates on parameter sets with $N_e^{(2)}\in(10,60)$: Larger values would imply pure small-field inflation in the observable range. Altogether, this means that the setting introduced here includes a transition from large-field inflation to small-field inflation after the modes constrained by CMB observations have left the horizon. This might cause a problem if it cannot be excluded that preheating sets in after the end of the first inflation, frustrating a subsequent restart of accelerated expansion. 

For this thesis it has been assumed that particle production is overcome by Hubble dilution, and inflaton decay around $\phi=0$ need not be accounted for. Taken into the extreme, one could consider the alternative of an uninterrupted inflation. However, while it proves difficult to rule out a viable scenario with one uninterrupted inflation, no parameter set has been found that allows for a smooth transition from large-field to small-field inflation such that both types of inflation with their typical features would show up in the observable range. Typically, inflation is either interrupted, where the interruption needs to take place outside the CMB range in order not to be ruled out by measurements; or inflation continues down to $\phi=0$ giving rise to many e-foldings for small $\phi$.  
The evolution either gets stuck around $\phi=0$ or there is at least such a strong expansion during this stage that there remains no possibility of observable large-field inflation.
This leads to the conclusion that parameter sets with $N_e^{(2)}\notin[10,60]$ should be discarded. Also within this range the resulting spectra coincide better with measurements for smaller $N_e^{(2)}$. 

It has been argued that the resulting spectra should not be altered significantly when the computation includes multifield dynamics. This is because multifield effects such as entropy modes are only expected when the path in field space is curved.    
This curvature, however, typically occurs only at one point in the evolution, where inflation is interrupted. For this paper both, multifield dynamics and possible preheating during the interruption are neglected. This should be treated more thoroughly in future work.

\section*{Acknowledgments}
Part of the calculations for this paper has been done with the free program LATTICEEASY for lattice simulations of scalar fields. We would like to thank the authors Gary Felder and Igor Tkachev.
We would like to thank Matthias Bartelmann, Jan M. Pawlowski and D\'{e}nes Sexty for useful discussions.
Simon Schettler acknowledges support by the IMPRS for Precision Tests of Fundamental Symmetries and the Institute for Theoretical Physics of Heidelberg University.


\end{document}